\documentclass[5p]{elsarticle}

\usepackage{graphicx}
\usepackage{amssymb}

\usepackage{lineno}

\journal{Astronomy and Computing}
\date{January 18, 2021}

\begin{document}

\begin{frontmatter}


\title{Development of a High Throughput Cloud-Based Data Pipeline for 21 cm Cosmology}



\author[UW]{Ruby Byrne}
\address[UW]{Physics Department, University of Washington, Seattle, Washington, United States}

\author[ASU]{Daniel Jacobs}
\address[ASU]{School of Earth and Space Exploration, Arizona State University, Tempe, Arizona, United States}

\begin{abstract}
We present a case study of a cloud-based computational workflow for processing large astronomical data sets from the Murchison Widefield Array (MWA) cosmology experiment. Cloud computing is well-suited to large-scale, episodic computation because it offers extreme scalability in a pay-for-use model. This facilitates fast turnaround times for testing computationally expensive analysis techniques. We describe how we have used the Amazon Web Services (AWS) cloud platform to efficiently and economically test and implement our data analysis pipeline. We discuss the challenges of working with the AWS spot market, which reduces costs at the expense of longer processing turnaround times, and we explore this tradeoff with a Monte Carlo simulation.
\end{abstract}


\end{frontmatter}


\section{Introduction}

The formation of the first stars and galaxies, and their later acceleration due to dark energy, can be probed by measuring the large-scale distribution of neutral hydrogen at high redshift (for reviews see \citet{Morales2010} and \citet{Liu2020}). Several arrays have been developed to measure the power spectrum of this cosmological signal, including the Low Frequency Array\footnote{\texttt{http://www.lofar.org}} (LOFAR; \citep{VanHaarlem2013}), the Donald C. Backer Precision Array for Probing the Epoch of Reionization\footnote{\texttt{http://eor.berkeley.edu}} (PAPER; \citep{Parsons2010}), the Murchison Widefield Array\footnote{\texttt{http://www.mwatelescope.org}} (MWA; \citep{Tingay2013}), and the Hydrogen Epoch of Reionization Array\footnote{\texttt{https://reionization.org}} (HERA; \citep{Deboer2017}). The hundreds of antennas and thousands of channels which give the needed sensitivity and redshift span generate a significant amount of data. For example, the MWA has generated a 28 PB data archive since it commenced in 2013.

Data analysis follows a traditional calibration and imaging approach, a compute intensive operation which is not trivially parallelizable. Analysis is further complicated by the need to distinguish foregrounds from the faint spectral signature of the cosmological background. This challenge emerges as a need to control for systematic error throughout the experiment and analysis to one part in 100,000; custom analysis codes are required to control for systematics in calibration, synthesis imaging, and error propagation. To date, all limits on the cosmological power spectrum have been limited by systematic biases that degrade measurement precision. These systematic floors are reached after processing hours or days of data. Each analysis iteration results in better identification of systematics and allows integration of more data for a deeper measurement. The iteration cycle is improved by testing on large amounts of data.

Recently, cloud computing has emerged as an alternative to traditional computing clusters for high-performance academic research computing, particularly of large astronomical data sets. \citet{Dodson2016} describes using the Amazon Web Services (AWS) cloud computing service to analyze the CHILES dataset, an example of parallelization used to process repeated measurements. \citet{Sabater2017} similarly calibrates LOFAR data with AWS. A related analysis, though not of radio astronomy data, was reported by \citet{Warren2016}, which describes processing satellite images in the cloud.

Cloud computing is particularly well-suited to episodic computation, where users require short periods of high computational throughput interspersed with periods of low usage. Dedicated clusters or small shared clusters can be expensive to maintain during periods of minimal usage and limited in their scalability during periods of heavy computation. The development of analysis techniques for radio cosmology measurements requires highly episodic computation as we identify systematics and test new analysis approaches on large data sets. The speed of this development cycle is limited by the testing turnaround time. 

Here we discuss how we have used cloud computing to routinely test analyses of data from the Murchison Widefield Array (MWA). We have used AWS to execute jobs in hundreds of parallel nodes, performing calibration, synthesis imaging, mosaicing, and power spectrum analysis on hundreds of TB of data. We describe our cloud pipeline and report finding on its efficiency, cost, and failure modes. We note that while the spot market mitigates costs, it extends testing turnaround times. To better understand this tradeoff we present a simple model that simulates the impact of the spot market on a typical analysis run. The simulation indicates that improvements in checkpointing and restart automation would offer faster overall execution time while retaining the spot market's cost savings.

\section{Background on Cloud Computing with AWS}
\label{S:2}

While a number of cloud computing platforms exist (Microsoft Azure, Google Cloud, etc.), this paper focuses on a workflow developed with AWS. We primarily use two AWS tools: Elastic Compute Cloud (EC2) for computation and Simple Storage Service (S3) for data storage. In this section we describe the basic functionality of these tools and define terminology used throughout the paper.

\subsection{EC2}

Cloud applications are run on instances that are started and stopped on command; costs are calculated in “instance-hours” and are incurred only from operating instances. This pay-for-use model is one of the primary advantages of cloud computing. AWS offers many instance types optimized for computation, memory, and storage. Instances are based on virtual central processing units (vCPUs), graphics processing units (GPUs), or both. The instance price, in dollars per instance-hour, reflects the size and capabilities of the instance. EC2 operates across 22 geographical regions that are further subdivided into 69 availability zones. Instance availability varies across regions.

EC2 instances are pre-configured with Amazon Machine Images (AMIs). While AWS provides templated AMIs, users can also build and save their own AMIs to produce customized instance environments. EC2 instances support many Linux and Windows operating systems.

Instances belong to one of two pricing models. On-demand instances have a fixed price that is consistent across regions. AWS service outages can, in rare cases, limit on-demand instance availability, but in general on-demand instances are available when requested and terminated by the user.

A cheaper, but less reliable, alternative to on-demand instances are spot instances. These typically cost a fraction of the on-demand price with the trade-off that they can be terminated by AWS at any time. The spot price operates as a market rate for a given instance type and increases during periods of high demand. If demand increases while the instance is running, AWS may terminate spot instances to increase capacity for the more expensive on-demand instance requests. Users can supply a maximum spot price when requesting spot instances. The instances will be terminated if the spot price exceeds this maximum price.

AWS overhauled its spot pricing system in 2018.\footnote{\texttt{https://aws.amazon.com/blogs/compute/ new-amazon-ec2-spot-pricing/}} Previously, spot instances were allocated based on a bidding system. Instance requests were fulfilled to the highest bids and the spot price was set to the highest unfulfilled bid. In 2018, in response to volatile spot pricing, AWS decoupled spot instance allocation from spot pricing. As a result, spot pricing is much more stable but is no longer a proxy for instance availability.

\subsection{S3}

The Simple Storage Service, or S3, is the most popular AWS system for persistent data storage and management. S3 is an object storage service rather than a file storage system. The top-level storage container is called a “bucket.” Buckets can be further subdivided by folders. S3 bucket access is customizable to support fully public, fully private, or read-only access. 

S3 storage has different classes with varying features and pricing. Two of the commonly used classes are S3 Standard and S3 Glacier. The Glacier class has a cost structure designed for long-term, low use, storage. It therefore has lower monthly storage costs but higher download costs, longer recall times, and deletion penalties if data are stored for less than a 90-day minimum storage duration.

\subsection{Interfacing}

There are a number of avenues for users to interact with AWS services. AWS provides a browser-based console\footnote{\texttt{https://aws.amazon.com/console/}} that allows users to review and manage cloud services. Command-line tools, such as the AWS command-line interface (CLI) package\footnote{\texttt{https://github.com/aws/aws-cli}}, enable users to download from or upload to S3 buckets. Users can access EC2 instances from the command line with SSH. Many users also use Application Programming Interfaces (APIs), such as boto,\footnote{\texttt{https://github.com/boto/boto3}} to interact with AWS cloud services.

\section{Data Processing}
\label{sec:pipeline}
We describe processing data from the MWA radio observatory. The MWA is an array of 128 stations, each comprising a grid of 16 dipole antennas phased to form a steerable 15 degree field-of-view. The interferometric output is a measure of the correlation between all pairs of stations, or baselines, as a function of frequency, polarization, and time. Data volumes therefore scale as the number of independent baselines, or $N(N+1)/2$ where $N$ denotes the number of stations, meaning that larger arrays produce substantially larger data volumes. Data are recorded continuously, integrated at a 2-second cadence, and divided into 2-minute observation files. The length of the 2-minute observations is chosen to be small compared to the amount of apparent sky rotation, enabling each observation to be imaged as stationary snapshots of the sky. Since first light in 2013, the MWA has recorded some 1.7 million of these files, amounting to 2 PB of data, from its cosmology program alone.

Data processing consists of two steps. First, correlated measurements, or visibilities, are gridded and Fourier transformed to produce an image of the sky at each time, frequency, and polarization. During this step, the 2-minute observation files can be processed independently in parallel. Second, repeated observations of the sky location are combined producing an ``image cube'' with two spatial axes and a frequency axis. We then estimate the cosmological power spectrum by Fourier transforming each axis of this cube and integrating in spherical shells.

Several pipelines have been developed to perform data analysis and power spectrum estimation. The U.S. team's analysis pipeline, described in depth by \citet{Barry2019a}, calibrates and images the data with the Fast Holographic Deconvolution (FHD) software package\footnote{\texttt{https://github.com/EoRImaging/FHD}} \citep{Sullivan2012} and perform precision power spectrum analysis with $\epsilon$PPSILON\footnote{\texttt{https://github.com/EoRImaging/eppsilon}} \citep{Jacobs2016}. The packages are open-source and written in the IDL programming language. Challenges addressed by this pipeline include precision imaging of diffuse structure, accounting for polarization with detailed instrument models, calibration of low level instrumental artifacts, and mitigation of gridding artifacts.

\begin{figure}
\centering
\includegraphics[width=0.9\columnwidth]{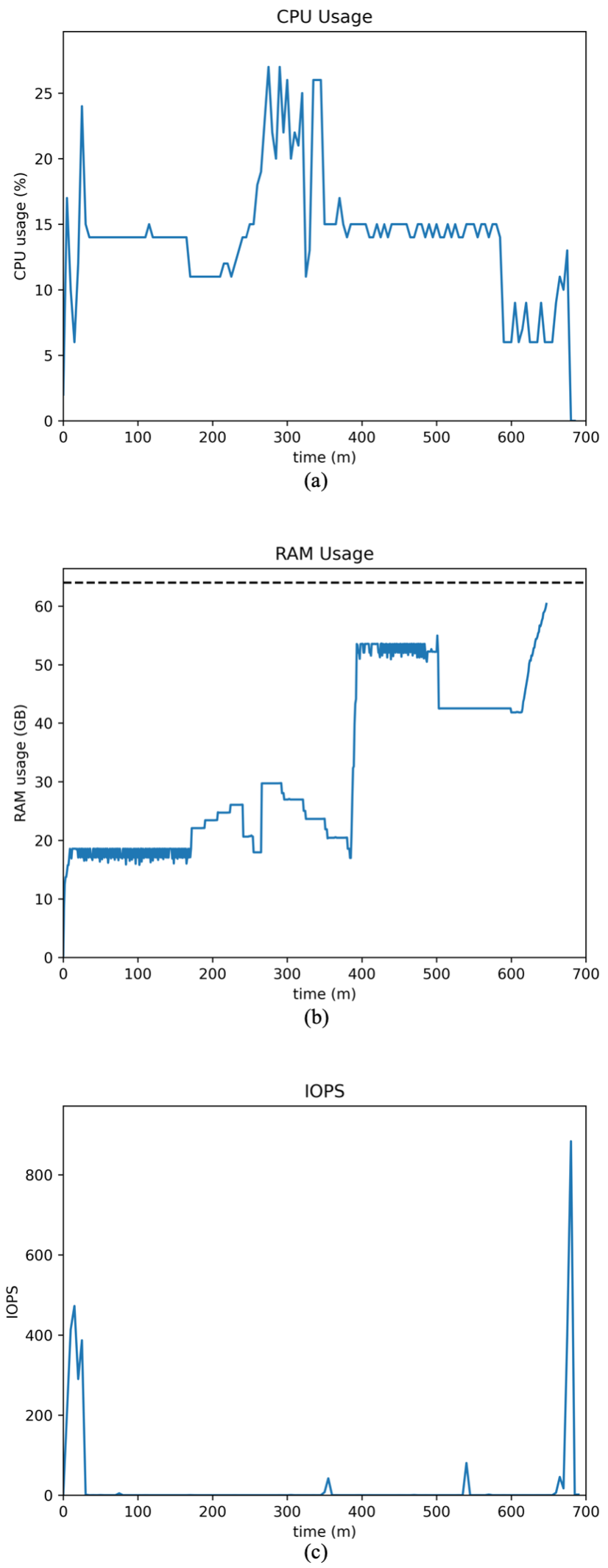}
\caption{Statistics from processing a 2-minute observation from the MWA with FHD. We used FHD's fully-polarized ``4-pol'' mode and implemented point source deconvolution. Processing was performed with an m5.4xlarge AWS instance. (a) gives the maximum CPU usage during one-minute time intervals throughout the run. (b) depicts the RAM usage, with the dotted black line representing the 64 GB RAM capacity of the m5.4xlarge instances. (c) gives the Input/Output Operations Per Second (IOPS).}
\label{fig:run_stats}
\end{figure}

In Figures \ref{fig:run_stats} we describe the processing of a single 2-minute observation with FHD in the cloud with AWS EC2. We use m5.4xlarge instances, which have 16 vCPUs and 64 GB of RAM, and we parallelize the job across all 16 vCPUs. FHD is a versatile software package; here we describe a particularly computationally-intensive processing run that includes calibration to the GLEAM catalog \citep{Hurley-Walker2017}, is fully Stokes polarized, and iteratively deconvolves compact foreground sources \citep{Sullivan2012}. Computational challenges include the initial calculation of model visibilities for 50,000 
sources, a task which scales linearly with sources, and the inclusion of full Stokes polarization modeling which models the correlation between all four polarization components requiring eight times the usual resources required by the usual analysis which neglects polarization. Figure \ref{fig:run_stats}a shows CPU usage, Figure \ref{fig:run_stats}b shows RAM usage, and Figure \ref{fig:run_stats}c shows the IOPS (disk reads and writes) for a job of this processing style.

On average, processing a single 2-minute observation in this way takes 633 minutes. At the time of writing, on-demand m5.4xlarge instances cost \$0.768 per instance-hour in the Virginia region, amounting to a total processing cost of \$8.10 per observation. Using spot instances reduces that cost to \$0.432 per instance-hour for the `a' availability zone, with spot pricing varying by up to \$0.1251 across availability zones. Processing an observation with a spot instances that does not experience spot termination costs about \$4.56. Using spot instances can save nearly half the cost over on-demand instances.\footnote{Since this analysis, we have transitioned to using the r5 class of instances. r5.2xlarge instances enable the style of processing depicted in Figure \ref{fig:run_stats} for an on-demand price of \$0.532 per instance-hour as compared to \$0.768 for m5.4xlarge instances. Users should regularly evaluate the available instance types to ensure that they choose most economical option.}

In \S\ref{s:workflow} we describe in detail the cloud workflow we developed to efficiently process many observations in parallel using virtual computing clusters.

\section{The AWS Cloud Workflow}
\label{s:workflow}

In this section we present the cloud-based data processing workflow we developed to process cosmological data sets from the MWA. The workflow supports high throughput parallelized processing of many observations. It is efficient, economical, and relatively simple to operate and train others to use. Since its development, we have trained five new users on the workflow. It is currently heavily used by three graduate students at the University of Washington. 

Developing this workflow consisted of two parts. First, we created an AMI for configuring EC2 instances to run our software pipeline. Next, we developed a auto-scaling cluster architecture.

\subsection{Developing the AMI}

We chose to store our instance configuration as an AMI. This has the advantage of being simple to create and use but is less flexible than other approaches because it can operate only on AWS instances. An extension of this work would be to use a lightweight and portable container such as a Docker container\footnote{\texttt{https://www.docker.com}} instead of an AMI. Containers would offer better stability across operating systems and allow for porting beyond AWS.

We configured an Ubuntu AMI to run our analysis software and to be compatible with ParallelCluster, our cluster management tool (see \S\ref{S:cluster_management}). We installed IDL with an appropriate licensing file and downloaded the IDL-based FHD and $\epsilon$PPSILON software packages from GitHub. We also installed Miniconda Python distribution and pyuvdata\footnote{\texttt{https://github.com/RadioAstronomySoftwareGroup/pyuvdata}}, a Python-based software package for interfacing with interferometric data sets. These are not required for our primary data analysis but expanded the usability of the AMI.

We used an inexpensive (m4.large) instance for software installation and path configuration. To test and debug our installation, we saved the AMI and used it to format a larger instance that allowed us to process a single test observation.

AMIs are stored in EC2 and can be copied between AWS accounts or downloaded. This enabled us to efficiently migrate between accounts without having to reconfigure the AMI.

We created a single AMI to serve most of the computing needs of our users. This AMI was adopted and further customized by different users within the collaboration. Those who made substantial changes saved their changes as additional AMIs. For example, one user, Michael Wilensky, adapted the workflow for running SSINS\footnote{\texttt{https://github.com/mwilensky768/SSINS}}, a Radio Frequency Interference (RFI) excision algorithm \citep{Wilensky2019}. He produced and saved a new AMI that includes the SSINS installation.

\subsection{Cluster Management}
\label{S:cluster_management}

In establishing our cloud-based high-performance computing (HPC) infrastructure, we sought tools that would provide the following elements:
\begin{enumerate}
\item A cluster interface such as a master instance equipped with a scheduler for submitting jobs.
\item A compute fleet for executing jobs, ideally one that is fully scalable to meet the needs of the queue while immediately terminating idle instances.
\item A shared file system or similar mechanism for communicating between the master and compute instances.
\end{enumerate}
We use the ParallelCluster open-source cluster management tool,\footnote{\texttt{https://github.com/aws/aws-parallelcluster}} which natively provides this functionality. ParallelCluster is developed by AWS and is open-source. It replaces the now-deprecated CFNCluster (we originally used CFNCluster and migrated to ParallelCluster after the latter's release).

ParallelCluster leverages the AWS CloudFormation and Auto Scaling Groups tools to create scalable high-performance computing (HPC) clusters.
It supports scheduling software that is commonly used with academic HPC clusters.
When we initially developed the cloud workflow, ParallelCluster supported four schedulers: Slurm, SGE, Torque, and a proprietary scheduler called AWS Batch. Our group had primarily used SGE on academic computing clusters, so we chose to retain that scheduler for our cloud-based processing. This facilitated our transition to the cloud by limited the amount of new code development required.

In May 2020, AWS announced that future releases of ParallelCluster will not support SGE or Torque schedulers on the basis that those open-source projects are not actively maintained.\footnote{\texttt{https://github.com/aws/aws-parallelcluster/wiki/ Deprecation-of-SGE-and-Torque-in-ParallelCluster}}. In response, we plan to migrate our cloud workflow to Slurm.\footnote{\texttt{https://github.com/EoRImaging/pipeline\_scripts/issues/15}}

ParallelCluster is launched from a local machine or dedicated instance. It produces an auto-scaling cloud cluster that is highly customizable with configurable keywords. These keywords specify, among other things: the operating system (we use Ubuntu 16.04, the latest Ubuntu operating system release supported by ParallelCluster); the instance type; the scheduling software (SGE in our case); the volume sizes for each the master and compute instances; the maximum number of compute instances in the cluster; the AMI (referenced with a unique AMI ID); the VPC; the instance pricing model (on-demand or spot); and, for spot instances, the maximum spot price. We also configure ParallelCluster with user tags, which are applied to all resources in the cluster and can be used to calculate the costs incurred by each user.

\subsection{The Cluster Workflow}

Under our workflow, users launch their own ParallelCluster cloud cluster with customized configuration settings. The users interact with the master instance with SSH. Software is installed on shared volumes, so code changes made on the master instance are applied across the cluster.

\begin{figure}
\centering
\includegraphics[width=\columnwidth]{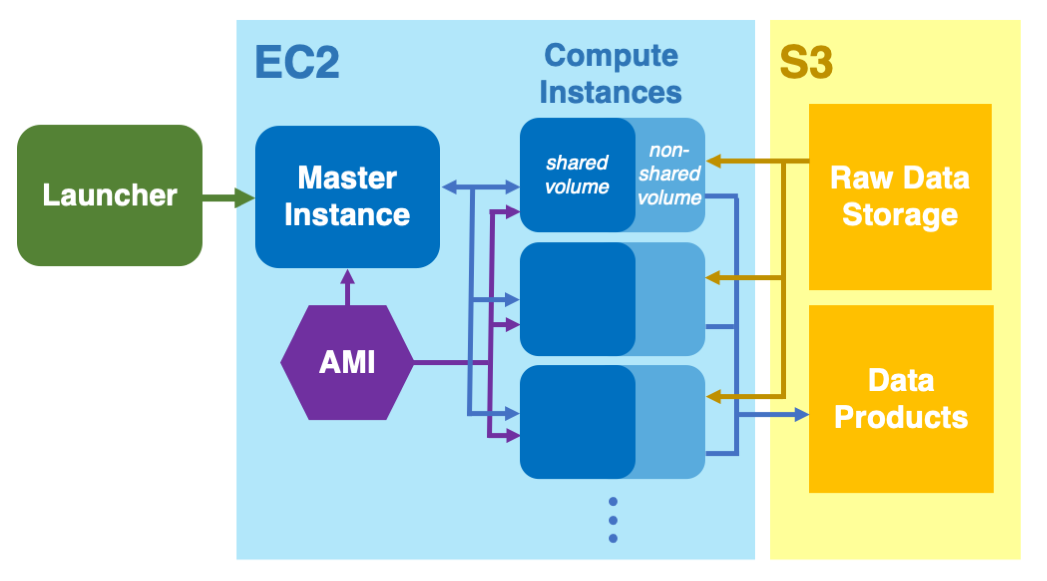}
\caption{Schematic of the data processing workflow with virtual supercomputing clusters on AWS. Users interact with the Master instance, which starts, stops, and assigns jobs to the Compute instances. Both the Master and Compute instances are modeled off the AMI, which stores the instance configuration and software installations. Master instances are on-demand; compute instances may be on-demand or spot. Raw data and data products are stored in S3 buckets.}
\label{fig:workflow}
\end{figure}

Users submit jobs to the SGE queue either individually (e.g. for a test of a single observation) or through a shell wrapper (e.g. for a processing run of many observations). The ParallelCluster software monitors the queue, assigns jobs to compute instances, and adjusts the size of the compute fleet up to the user-defined maximum queue size. Jobs consist of a top-level shell wrapper that downloads raw data from S3, runs the IDL scripts for data processing, and uploads the data products and output logs to S3. Background processes back up data products to S3 every 30 minutes in case of an unexpected instance termination. They also monitor spot instance termination notices, which go into effect two minutes before termination, to ensure that data products and output logs are not lost when spot instances are terminated from over-demand. All wrappers are available open-source.\footnote{\texttt{https://github.com/EoRImaging/pipeline\_scripts}} See Figure \ref{fig:workflow} for a graphical representation of this workflow.

\section{Working with the Spot Market}
\label{sec:spot_market}

The AWS spot market allows users to trade reliability for cost savings: EC2 instances can be purchased at steeply reduced cost if the user can tolerate some probability of unexpected instance termination. There is extensive literature exploring the potential savings from using spot instances \citep{Yi2010, Javadi2011, Jung2011, Mazzucco2011, Voorsluys2012, AgmonBen-Yehuda2013, Poola2014, He2015, Karunakaran2015, Haghshenas2019}.

Spot market performance is highly variable and difficult to predict \citep{Javadi2011, AgmonBen-Yehuda2013}. Since 2018, when AWS decoupled spot pricing and instance termination behavior, users cannot use spot instance pricing history to estimate termination rates \citep{Khandelwal2018}. The lack of transparency and high volatility of the spot market imposes barriers to evaluating its effectiveness in mitigating computational costs. Low spot termination rates can deliver exceptional cost savings while only marginally increasing total processing time, while high spot termination rates lead to much longer processing turnaround times with little to no cost savings.

As an illustration of the spot market's volatility, we discuss a case study of two large data processing runs. In early February 2020, we experienced relatively good spot market performance during processing of a set of 103 observations. The data were processed in the full polarization and deconvolution mode described in \S\ref{sec:pipeline}. 91 of the 103 observations were processed to completion in the inital run; 12 were spot terminated midway through processing. To complete the processing run, we resubmitted the 12 spot terminated jobs. All 12 jobs successfully ran to completion. The run of $\sim10$ compute hours completed in $\sim24$ wall clock hours with an effective termination rate of 10\% per job. From the AWS Cost Explorer tool we estimate that this processing run cost \$461.
Later that month we discovered an error in the normalization of the images used in the deconvolution step of FHD's processing. We documented the bug in a GitHub issue,\footnote{\texttt{https://github.com/EoRImaging/FHD/issues/198}} resolved it with a pull request,\footnote{\texttt{https://github.com/EoRImaging/FHD/pull/199}} and initiated data reprocessing.

This time, spot instance terminations affected 37 of the 99\footnote{Upon inspection of the 103 jobs in the first round, we found that four observations had low data quality. We expect that this stems from poor calibration performance, potentially resulting from residual RFI contamination that evaded flagging in our data pre-processing steps. It could also be exacerbated by high ionospheric activity, as refraction through the ionosphere can contaminate the images and reduce agreement with the calibration model \citep{Loi2015, Jordan2017}.} observations processed. Upon resubmitting those jobs, only half ran to completion. 18 of the 37 jobs were spot terminated. We resubmitted the terminated jobs a second time and 13 of the 18 jobs were terminated. Next, 7 of the 13 jobs were terminated. Then, 1 of the 7 jobs was terminated. We resubmitted that job only to have it spot terminated again. Finally, we submitted it and got it to run to completion. In all, we submitted jobs 7 times to achieve successful analysis of the 99 observations. The effective spot termination rate was 44\% per job and the approximate cost was \$823. This time the $\sim10$ compute-hour run took almost a week to complete.
The second run took roughly four times longer than the first at nearly double the cost. It is clear that spot market performance has major implications for the cost and time efficiency of large data processing runs. 

\begin{figure}
    \centering
    \includegraphics[width=\columnwidth]{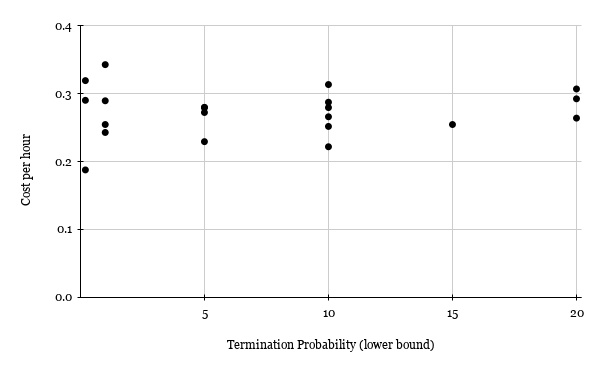}
    \caption{Like fine wines, instance pricing is not indicative of quality. Here we plot the spot market price for the m5.4xlarge instances versus the AWS Instance Advisor ``frequency of interruption'' metric. This metric is reported as range, and we plot the lower bound. Note that the highest bin, 20\%, is a lower bound including all fractions up to 100\%. Data captured on Aug 27, 2020 from AWS Instance Advisor.}
    \label{fig:price_tr}
\end{figure}

Since 2018, the only documentation on spot termination rates provided by AWS is the Instance Advisor tool,\footnote{\texttt{https://aws.amazon.com/ec2/spot/instance-advisor/}} which lists coarse ``frequency of interruption'' ranges from $``<5\%$'' to ``$>20\%$.'' These metrics can be useful for making rough comparisons between difference instances types (and it is interesting to note that they are not correlated with spot prices --- see Figure \ref{fig:price_tr}). However, they are wholly inadequate for predicting spot market performance during our processing runs for three key reasons. First, the Instance Advisor metric ranges are too coarse. Below 20\%, the metric bins span 5\%. However, the maximum metric category is ``$>20\%$,'' including values of up to 100\%. Next, the Instance Advisor metric does not capture time variations in spot termination rates. We have found that spot performance is highly variable over timescales of days or weeks, and this is not captured by the Instance Advisor tool.

\begin{figure}
    \centering
    \includegraphics[width=\columnwidth]{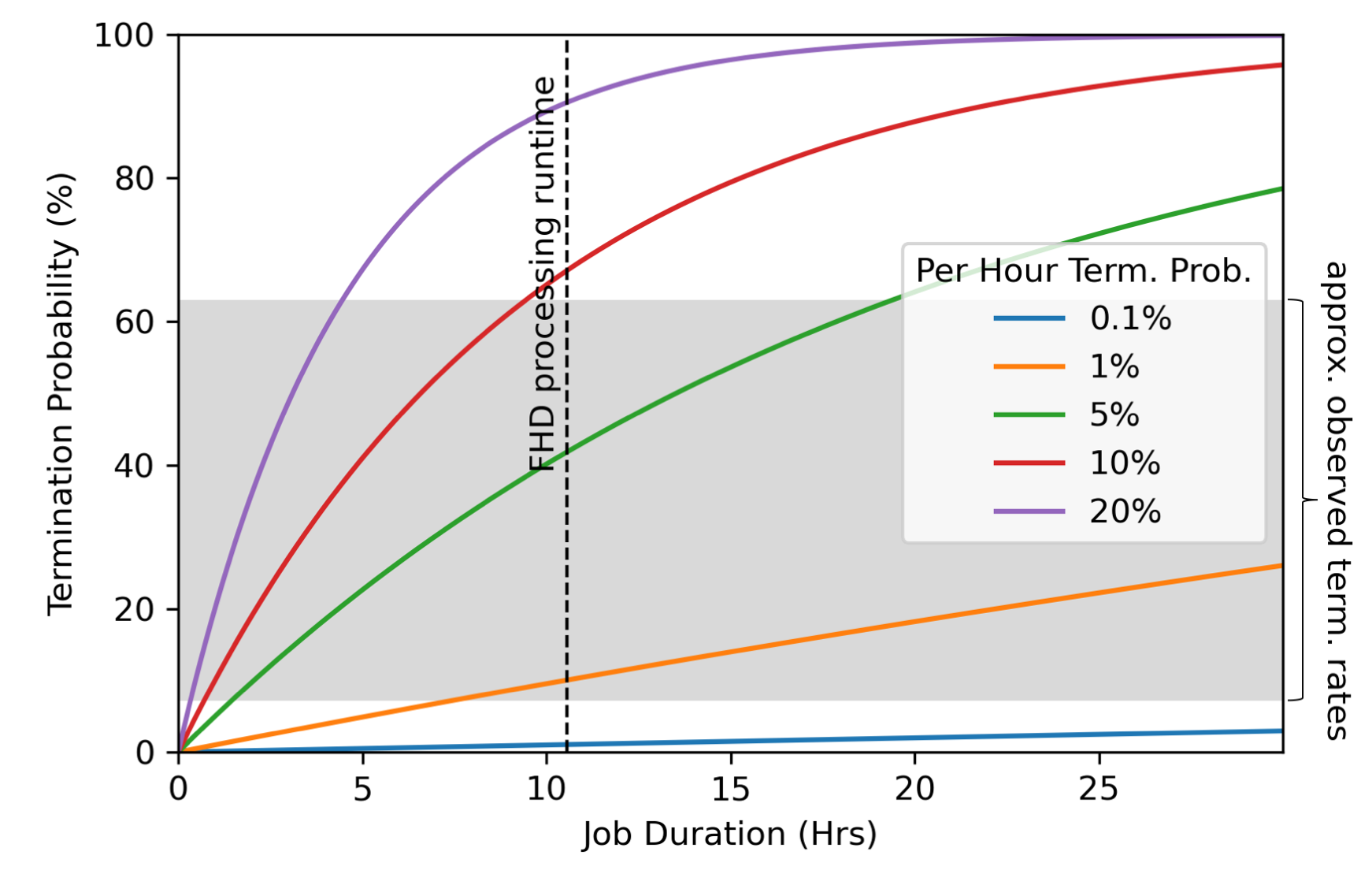}
    \caption{
    The AWS Instance Advisor metric has an ill-defined relationship to the probability that a particular job will be terminated. If we assume that spot termination rates (in \% per hour) are constant over job duration, the total termination probability follows an exponential law. We therefore see that the spot termination rates' correspondence to a job's termination probability is highly dependent on the job's duration. 
    The vertical dashed line in this plot marks the 633-minute runtime of our job. The shaded grey region spans the observed termination rates from two large processing runs in February 2020. Their intersection indicates a range of underlying hourly termination rates between .7\% and 9\%/hour.}
    \label{fig:term_prob}
\end{figure}

Finally, the units of the Instance Advisor metric are not well-defined. According to AWS, the Instance Advisor metric ``represents the rate at which Spot has reclaimed capacity during the trailing month.'' However, we expect (and observe in practice) that longer-running jobs have a greater probability of termination than shorter jobs (Figure \ref{fig:term_prob}). This begs the question, what is the characteristic runtime for jobs contributing to the Instance Advisor metric? Without this additional information, we cannot know if our jobs are representative of those terminated at the rate given by the Instance Advisor. There is no clear way to map the Instance Advisor metric to the expected termination rate for our data processing jobs. 

Noting the inadequacy of the Instance Advisor metric for meaningfully predicting spot market behavior, we turn to a simplified Monte Carlo simulation. The simulation explores the relationships between spot termination rates, cost, total processing time, and the time between save points in the analysis pipeline.

We simulate a processing run with 100 jobs that complete in 633 minutes and run on a virtual cluster with the capacity to run all 100 jobs in parallel. We further allow that each job takes 2 minutes to start and stop; this can account for the time spent copying data from S3 and uploading pipeline outputs. This simulation has 1-minute time resolution. We explore a variable number of equally-spaced checkpoints throughout the jobs that represent saving the full job state. The simulation assumes a constant spot termination probability over the timescale of the run. In practice we observe variable spot termination rates, and job terminations are a stochastic process subject to uncertain measurement. However, the approximation that the termination probability is constant for the duration of a run is a useful simplifying assumption, and more data are needed to justify a more complex model. We explore rates of 0-15\% termination probability per hour. On the upper end of this range, a termination rate of 15\% termination probability per hour corresponds to a 82\% probability that a single 633-minute job will be spot terminated (see Figure \ref{fig:term_prob}).

\begin{figure}
\centering
\includegraphics[width=\columnwidth]{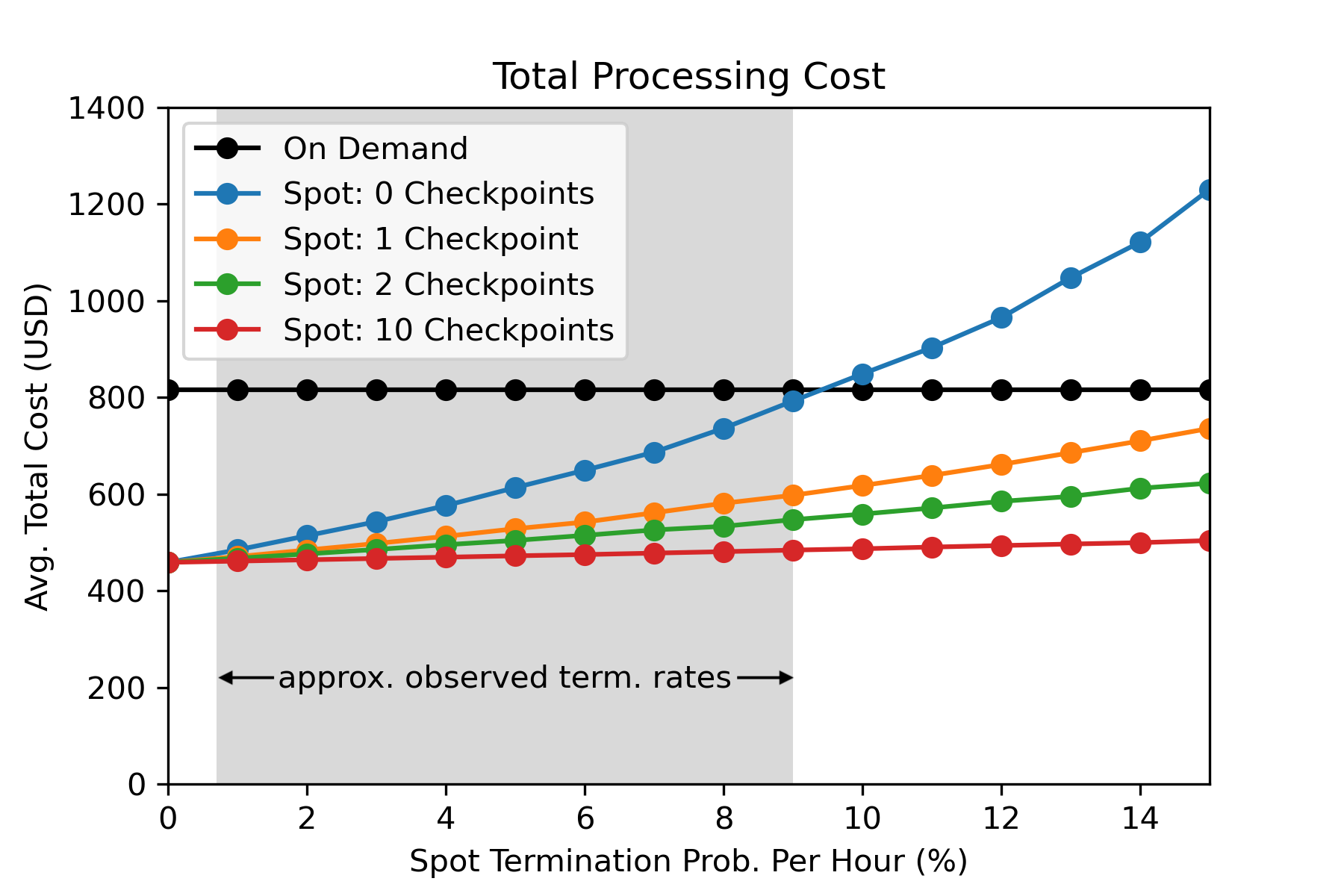}
\caption{Lost compute time due to spot terminations adds cost. Here we use a Monte Carlo simulation to calculate the average cost of a 100-job data processing run. Each job has a total runtime of $\sim10$ hours (633 minutes). We explore variable spot termination rates and note that very high termination rates can push the total cost beyond that of on-demand computing. We find that frequent checkpointing mitigates costs by reducing rerun times. The shaded grey region spans the approximate spot termination probabilities observed over two large processing runs in February 2020.}
\label{fig:sim_cost}
\end{figure}

In Figure \ref{fig:sim_cost} we examine the cost associated with processing 100 jobs. Higher spot termination rates lead to higher overall costs due to the extra runtime required to re-run spot-terminated jobs. For high spot termination rates, the total cost could even exceed the on-demand cost. Frequent checkpointing mitigates rerun costs by allowing restarted jobs to commence from an intermediate point.

\begin{figure}
\centering
\includegraphics[width=\columnwidth]{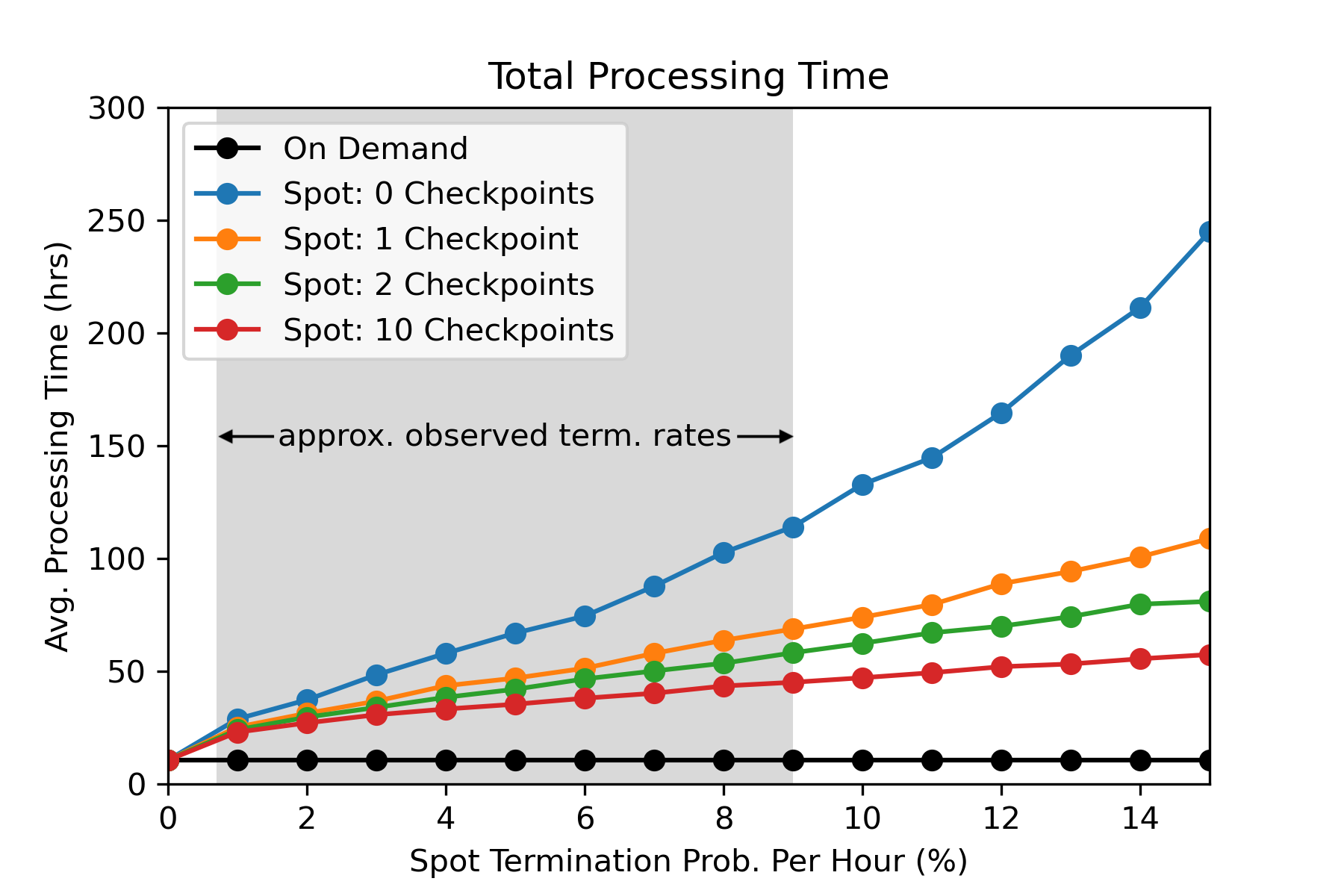}
\caption{Spot terminations increase processing turnaround times. Here we simulate processing 100 observations with spot and on-demand instances. Although the simulated cloud cluster can run all 100 jobs in parallel, limiting the total processing time to a single job's $\sim10$-hour runtime, spot terminations increase the processing time days or weeks. The shaded grey region in this plot spans the approximate spot termination probabilities observed over two large processing runs in February 2020.}
\label{fig:sim_time}
\end{figure}

We also explore the total time required to process the 100 jobs (Figure \ref{fig:sim_time}). We simulate that spot terminated jobs are restarted 4 hours after the completion (or termination) of the last running job. It is clear that spot terminations can significantly expand the total processing time. Even when all jobs can run in parallel, spot terminations necessitate re-runs that can make the total processing time much longer than a single job's runtime. For time-sensitive processing, it may be infeasible to budget this additional time. Using spot instances with high spot termination rates can substantially slow research development, potentially with minimal cost savings over on-demand instances, as shown in Figure \ref{fig:sim_cost}. Again, frequent checkpointing mitigates this effect by shortening the runtime of resubmitted jobs.

Data processing turnaround time could be improved by automatically resubmitting terminated jobs. Our simulation's 4-hour lag before job resubmission is a realistic estimate when jobs are manually resubmitted. We could instead capture spot termination notices and use them to automatically restart jobs. However, such automation makes assumptions about the cause of spot terminations and runs the risk of entering a loop. Because the spot market is highly variable, waiting to resubmit jobs allows the underlying reason for termination to abate. We have considered automated job restarting, potentially calibrated against a market model, as a future extension to our workflow.

Furthermore, we have seen some evidence that large processing runs can drive the spot market: spot termination rates increase when we run many spot instances at once. Although this is difficult to document, we have had some success reducing spot terminations by limiting the maximum cluster size. Instead of running all 100 jobs at once, we might limit our virtual cluster to only process 50 in parallel. The total processing time would then be at minimum twice the processing time associated with using on-demand instances and likely much longer than that due to spot terminations.

As shown in the simulation presented in Figures \ref{fig:sim_cost} and \ref{fig:sim_time}, the best antidote to the looming threat of spot terminations is frequent checkpointing \citep{Yi2010, Jung2011}. Upon termination, each job can resume processing at the last checkpoint achieved. 
FHD has built-in checkpointing, saving the results of particularly computationally expensive steps including visibility gridding, source modeling, calibration, and deconvolution. Our cloud workflow includes a parallel process which backs up FHD outputs to S3 every half hour. Furthermore, we capture spot termination notices and upload outputs to S3 before instance termination. Despite the twice hourly backups, outputs are generated at junctions between pipeline elements that can take hours to run. As visualized in Figure \ref{fig:run_stats}c, the intervals between FHD checkpoints range from just over an hour to about 5 hours. Furthermore, FHD uses incomplete checkpointing. Some intermediate data products are simply too time-intensive to write in their entirety and must be recalculated upon restart. In practice, terminations have been observed to significantly extend FHD runtimes. This effect could be mitigated by adding more checkpointing to the FHD pipeline, provided that the time required to write intermediate data products does not significantly impact the overall runtime.

We can use our simulation to estimate the spot termination rate of observed processing runs. If we approximate FHD checkpointing as two evenly-spaced checkpoints (denoted in green in Figures \ref{fig:sim_cost} and \ref{fig:sim_time}), then we
can match the number of reruns required in practice to those expected in simulation. Based on this assumption, our processing run in early February that experienced a 10\% job termination rate and required one rerun had approximately a 0.7\% termination probability in any given hour. The next data processing run, that experienced a 44\% job termination rate and required six reruns to complete, had approximately a 9\% termination probability in any given hour.


\section{Discussion}
\label{s:discussion}

Cloud computing has matured within the last decade, and Infrastructure as a Service (IaaS) has taken root in everyday technologies. In the lifetime of this project, which began in 2015, the scope and complexity of offerings has grown exponentially. Much of this development was driven by commercial needs; cloud computing tools for academic research have lagged behind private-sector advancement. Tools such as AWS's ParallelCluster have brought much-needed new investment to academic areas. 

Even so, migrating to the cloud poses some challenges to academic research groups. There can be a steep learning curve for effectively navigating cloud-based computing tools. AWS documentation can be thin, and existing documentation is rarely geared towards academic researchers. Advising from AWS technical experts at the University of Washington e-Science Institute proved invaluable to this work. As documentation expands and cloud computing becomes a more integral part of researchers' toolkits, we expect easier and more efficient development of cloud-based data processing pipelines.

One predominant challenge to using AWS is the difficulty of predicting costs. Although the AWS Cost Explorer tool catalogs expenditures, and tagging resources can link costs with individual users, estimating costs of a planned processing run can be difficult and time-consuming. Some costs can be unexpected for users who have not studied AWS documentation in detail, and there is no mechanism to cap costs to prevent overruns.

One early instance of unexpected costs occurred during ingest of $\sim$200TB of raw data into S3. The data were downloaded from the MWA archive to an EC2 instance, converted from raw binary to the FHD-readable uvfits file format, and uploaded to S3. We realized that instance storage was imposing a bottleneck and changed the default EC2 storage type to a solid state disk. This resolved the bottleneck and increased the data rate substantially. It also also pushed the IOPS into a non-free tier; in this way we accumulated some \$3,000 of unexpected charges. 

Later, we incurred a smaller unexpected charge when we tried to reduce data product storage costs by moving a data set to Glacier. Soon after, we determined that we could safely delete a subset of that data and did so, not realizing that AWS charges a penalty for early deletion of Glacier data. Because the data had been stored on Glacier for less than 90 days, we incurred a fee of several hundred dollars. 

Our largest accidental charge resulted from a security breach. AWS manages account access through Access Keys, which must be securely stored (and ideally changed frequently) to prevent account breaches. Our account security was compromised when a user accidentally published their Access Key to a public GitHub repository. An unauthorized user located the key and gained access to the account, where they initiated hundreds of EC2 instances. In the hours before we identified the breach and deleted the compromised Access Key, the account incurred over \$6,000 in fraudulent EC2 charges. Security breaches such as this one are easily preventable if users adopt good security practices, but it nonetheless underscores a security vulnerability unique to cloud computing. 

These charges might have been expected by an experienced AWS user; we were new to cloud computing and discovered pitfalls through trial and error. For small research groups with tight budgets, unexpected cost overruns can pose an insurmountable barrier to leveraging cloud computing resources.

With policies routinely changing and new services being added, translation of rules and rates into a cost estimate presents a challenge not unlike a tax filing. The typical method for discovering actual costs is experimental, a process requiring capital. For this reason AWS has had occasion to provide credits to academics for the purposes of experimentation. Since many academics, will, like us, find themselves needing to associate a personal credit card with the AWS account, large unforgiven overages presents a risk. In the large errors Amazon did eventually forgive the charges, however not without first threatening to send bills on our personal credit to collections.

The spot market offers a simple cost mitigation method, as spot instances are generally much cheaper than their on-demand counterparts. However, as we discuss in \S\ref{sec:spot_market}, high spot termination rates and highly variable spot market behavior can substantially extend processing times and erodes the potential savings. The spot market is best suited to pipelines with significant checkpointing, short job runtimes, or both. Time-sensitive processing runs benefit from using on-demand instances.

The pay-for-use model also represents a paradigm shift for researchers accustomed to using dedicated academic clusters. If analyses are expected to be tested in the cloud, users must be empowered to spend resources on unsuccessful runs. A group must dedicate ample funding to computation for testing and exploration. We found that new cloud users were consistently hesitant to use pay-for-use resources and apologetic when unsuccessful runs ``wasted'' money. This can slow research progress.

Overall, for our use case, under our unique circumstances of location and timing, the benefits of cloud computing with AWS outweighed the challenges. That said, the time investment required to establish a cloud-based pipeline and the ongoing costs of funding a pay-for-use computing model mean that academic research groups should plan for the unique development and management costs of migrating their computation to the cloud followed by a sustained effort to stay on top of changing market rules and conditions.

\section{Conclusion}

Using the AWS cloud computing platform, we have produced an efficient processing workflow for radio cosmology data. Our workflow is highly scalable, which permits faster testing turnaround times than with typical academic computing clusters. This enables rapid development of the novel analysis techniques needed to mitigate systematics in our data processing pipeline.

We note that substituting spot instances for on-demand instances can reduce computational costs at the expense of longer processing turnaround times. Our simplified Monte Carlo simulation, presented in \S\ref{sec:spot_market}, provides a metric for evaluating this tradeoff. Addition of checkpointing and automated job resubmission could enable more efficient spot market utilization, though this development runs the risk of designing too specifically to the artificial conditions of the spot market which are likely to continue evolving. Additionally, we have encountered evidence that large processing runs can drive the spot market. Spot market performance is therefore improved by using small clusters. This limits the cloud clusters' scalability and eliminates one of the predominant advantages of cloud computing.

In \S\ref{s:discussion} we discuss some practical challenges associated with migrating academic research computing to the AWS cloud. High-performance cloud computing requires skills that do not always overlap with academic research training. The pay-for-use model attaches a price tag to trial-and-error exploration that can be daunting and costly for academic researchers. We anticipate that, in the coming years, cloud technologies will become ever-more accessible, and we hope that case studies like the one presented here can help researchers use cloud computing more effectively.

As cloud technologies continue to mature, they will take on an ever more integral role in academic research computing. In particular, observational cosmology research necessitates enormous data sets and computationally expensive processing. The next generation of experiments boast larger arrays that can achieve improve sensitivity at the cost of larger data volumes. We show that these styles of analysis can benefit from cloud computing technologies, and we expect that next-generation cloud computing tools will further facilitate astronomical data analysis in the cloud.

\section*{Acknowledgemets}

We thank Nichole Barry, Jon Ringuette, and Michael Wilensky for their contributions to the AWS cloud workflow. Computation on AWS was supported in part by the University of Washington student-led Research Computing Club with funding provided by the University of Washington Student Technology Fee Committee. This project was made possible by computing credits from the Amazon/Square Kilometer Array Astrocompute initiative. We thank Lori Clithero, Aaron Bucher, and Sean Smith of AWS and Rob Fatland and Amanda Tan of the University of Washington e-Science Institute for their support and technical guidance. Finally, many thanks to Karen Jacobs for manually scraping AWS Instance Advisor data and Kyle Aitken for the workflow schematic in Figure \ref{fig:workflow}. This work was directly supported by NSF grants AST-1613855, 1506024, 1643011, and 1835421.





\bibliographystyle{model2-names}
\bibliography{cloudlib.bib}







\end{document}